\documentclass{book}
\usepackage[contrib,lang=british]{ems-book} 

\usepackage[square,numbers]{natbib}

\newcommand{\N}{{\mathbb{N}}}

\newcommand{\di}{{\rm d}}

\newcommand{\fv}{{\bf f}}

\newcommand{\rv}{{\bf r}}

\newcommand{\itTh}{{\mathit \Theta}}

\newcommand{\bmath}{\begin{eqnarray}}
\newcommand{\emath}{\end{eqnarray}}

\newcommand{\de}{\delta}

\begin{document}
\mainmatter

\title{THE LIEB-OXFORD BOUND AND THE OPTIMAL TRANSPORT LIMIT OF DFT}
\titlemark{LIEB-OXFORD BOUND AND SCE}

\emsauthor{1}{Michael Seidl}{M.~Seidl}
\emsauthor{2}{Tarik Benyahia}{T.~Benyahia}

\emsauthor{3}{Derk P. Kooi}{D.P.~Kooi}

\emsauthor{4}{Paola Gori-Giorgi}{P.~Gori-Giorgi}


\emsaffil{1}{Department of Chemistry \& Pharmaceutical Sciences and Amsterdam Institute of Molecular and Life Sciences (AIMMS), Faculty of Science, Vrije Universiteit, De Boelelaan 1083, 1081HV Amsterdam, The Netherlands }
\emsaffil{2}{Department of Chemistry \& Pharmaceutical Sciences and Amsterdam Institute of Molecular and Life Sciences (AIMMS), Faculty of Science, Vrije Universiteit, De Boelelaan 1083, 1081HV Amsterdam, The Netherlands }
\emsaffil{3}{Department of Chemistry \& Pharmaceutical Sciences and Amsterdam Institute of Molecular and Life Sciences (AIMMS), Faculty of Science, Vrije Universiteit, De Boelelaan 1083, 1081HV Amsterdam, The Netherlands }
\emsaffil{4}{Department of Chemistry \& Pharmaceutical Sciences and Amsterdam Institute of Molecular and Life Sciences (AIMMS), Faculty of Science, Vrije Universiteit, De Boelelaan 1083, 1081HV Amsterdam, The Netherlands \email{p.gorigiorgi@vu.nl}}

\classification[YYyYY]{XXxXX}


\begin{abstract}
We review and illustrate with several examples the connection between the search for lower bounds for the optimal constant in the Lieb-Oxford inequality and the optimal transport limit (or strictly-correlated-electrons limit) of density functional theory. We focus in particular on several contributions from Elliott Lieb which already hinted at this connection.
\end{abstract}

\makecontribtitle


\section{Introduction}
Elliott Lieb's work has been a continuous source of inspiration for researchers working on density functional theory (DFT) and quantum chemistry in general. In this small chapter we have chosen to focus on three among Elliott's papers (denoted throughout as L79 \cite{Lie-PLA-79}, LO81 \cite{LieOxf-IJQC-81} and L83 \cite{Lie-IJQC-83}), which had a profound influence on our work on the so-called strictly correlated electrons (SCE) limit of DFT (also known as the optimal transport or semiclassical limit of the Levy-Lieb functional), and its applications to the Lieb-Oxford inequality. We will go through a series of simple worked-out SCE examples that illustrate and support specific sentences and equations in these three papers. We hope that, despite their simplicity, Elliott will enjoy these examples that have been crafted here explicitly for him.

The chapter is organised as follows. We will first review, in Sec.~\ref{sec:LO}, the Lieb-Oxford inequality  \cite{Lie-PLA-79,LieOxf-IJQC-81}, including lower bounds on the optimal constant for different particle numbers $N$ derived from SCE calculations. The theory and ideas behind SCE are illustrated and connected to Elliott's work in Sec.~\ref{secSCE}, with various worked-out examples, including (Sec.~\ref{sec:LOSCE}) a construction from LO81 \cite{LieOxf-IJQC-81}, which was essentially a SCE state (although formulated at the time in a slightly different way), discussing a simple generalization. Then in Sec.~\ref{secBumps} we use SCE to do a computation that was imagined by Elliott in L83 \cite{Lie-IJQC-83}. Finally, we conclude with what we believe is an intriguing question on the optimal constants in the LO inequality for different particle numbers $N$.

\section{Lieb-Oxford (LO) bound}\label{sec:LO}
\subsection{General concepts}
Any (correctly normalized and antisymmetrized) $N$-electron wave function $\Psi$ is mapped onto an electron density
$\rho(\rv)\equiv\rho_{\Psi}(\rv)$, which we write $\Psi\mapsto\rho$, by the usual definition
\bmath
\rho(\rv)\;=\;N\sum_{\sigma_1,...,\sigma_N}\int \mathrm{d} \rv_2 \cdots\int \mathrm{d} \rv_N\,
\Big|\Psi\big(\rv,\sigma_1;\,\rv_2,\sigma_2;\,...;\,\rv_N,\sigma_N\big)\Big|^2.
\emath
The density is normalized to $N$, the total number of electrons in the system,
\bmath
\int d^3r\,\rho(\rv)\;=\;N.
\emath
In terms of the operator
\bmath
\hat{V}_{\rm ee}\;=\;\frac12\sum_{i,j=1}^{N\,(i\ne j)}\frac1{|\rv_i-\rv_j|},
\emath
the Coulomb electron-electron repulsive energy in the state $\Psi$ is the expectation
\bmath
I(\Psi)\;=\;\langle\Psi|\hat{V}_{\rm ee}|\Psi\rangle,
\emath
using the notation $I(\Psi)$ of L83, see eq (5.3) there. The Hartree functional
\bmath
U[\rho]\;=\;\frac12\int d^3r\int d^3r'\,\frac{\rho(\rv)\,\rho(\rv')}{|\rv-\rv'|},
\emath
denoted as $U[\rho]\equiv D(\rho)$ in L83, is called the direct part of $I(\Psi)$, while its indirect part is the difference
\bmath
W[\Psi]\;=\;\langle\Psi|\hat{V}_{\rm ee}|\Psi\rangle\,-\,U[\rho_{\Psi}].
\emath
In the notation of L83, see Eq.~(5.4) there, this same definition reads
\bmath
E(\Psi)\;=\;I(\Psi)\,-\,D(\rho_{\Psi}).
\emath
For clarity, we report in table~\ref{tab:notation} an overview of the notation used in LO81 and our previous work (denoted SVG) on the search of optimal constants for the bound  \cite{SeiVucGor-MP-16}.

\begin{table}
\begin{tabular}{|l|c|c|}\hline
Quantity & L83 & SVG \\\hline\hline
Coulomb repulsive energy                    & $I(\Psi)$ & $\langle\Psi|\hat{V}_{\rm ee}|\Psi\rangle$ \\\hline
direct   part of $I(\Psi)$ (Hartree energy) & $D(\rho)$ & $U[\rho]$ \\\hline
indirect part of $I(\Psi)$                  & $E(\Psi)$ & $W[\Psi]$ \\\hline
\end{tabular}
\caption{Notation used in L83 \cite{Lie-IJQC-83} and in SVG \cite{SeiVucGor-MP-16}.}
\label{tab:notation}
\end{table}

\subsection{The bound}\label{secTheBound}
Referred to as the Lieb-Oxford (LO) bound in the literature, Eq.~(6) of LO81 or Eq.~(5.7) of L83 reads
\bmath
\langle\Psi|\hat{V}_{\rm ee}|\Psi\rangle\;\ge\;U[\rho_{\Psi}]\,-\,C\int \mathrm{d} \rv\,\rho_{\Psi}(\rv)^{4/3},
\label{LObound_Original}\emath
where the constant $C>0$ has an unknown optimum, defined as the minimum value for which the inequality holds for any possible system. Writing
\bmath
C\;\ge\;\frac{U[\rho_{\Psi}]-\langle\Psi|\hat{V}_{\rm ee}|\Psi\rangle}{\int \mathrm{d} \rv\,\rho_{\Psi}(\rv)^{4/3}}\;\equiv\;\lambda_C[\Psi]
\label{LObound_Modify}\emath
(alternatively, SVG uses $\lambda[\Psi]=A_3\,\lambda_C[\Psi]$, where $A_3=\frac34(\frac3{\pi})^{1/3}\approx0.739$), we see that this optimum is a maximum (or a supremum),
\bmath
C\;=\;\sup_{\Psi}\,\lambda_C[\Psi].
\label{CsupPsi}\emath
We can also restrict ourselves to wave functions $\Psi\mapsto N$ with a given particle number $N$, and define
\bmath
C(N)\;=\;\sup_{\Psi\,\mapsto N}\,\lambda_C[\Psi]\qquad\qquad(N\in\N). \label{eq:CNdef}
\emath
Lieb and Oxford have then shown in LO81 that the corresponding unknown optimal constants $C(N)$ are monotonically increasing with growing $N$,
\bmath
C(N)\;\le\;C(N+1),\qquad\qquad\lim_{N\to\infty}C(N)=C.
\label{eq:monotCN}
\emath
The LO bound has played (and continues to play) a very important role in the development of approximate exchange-correlation functionals in DFT, as very clearly reviewed and illustrated in the Chapter by Perdew and Sun in this same volume. 

\begin{figure}[htb]
\includegraphics[width=0.6\columnwidth]{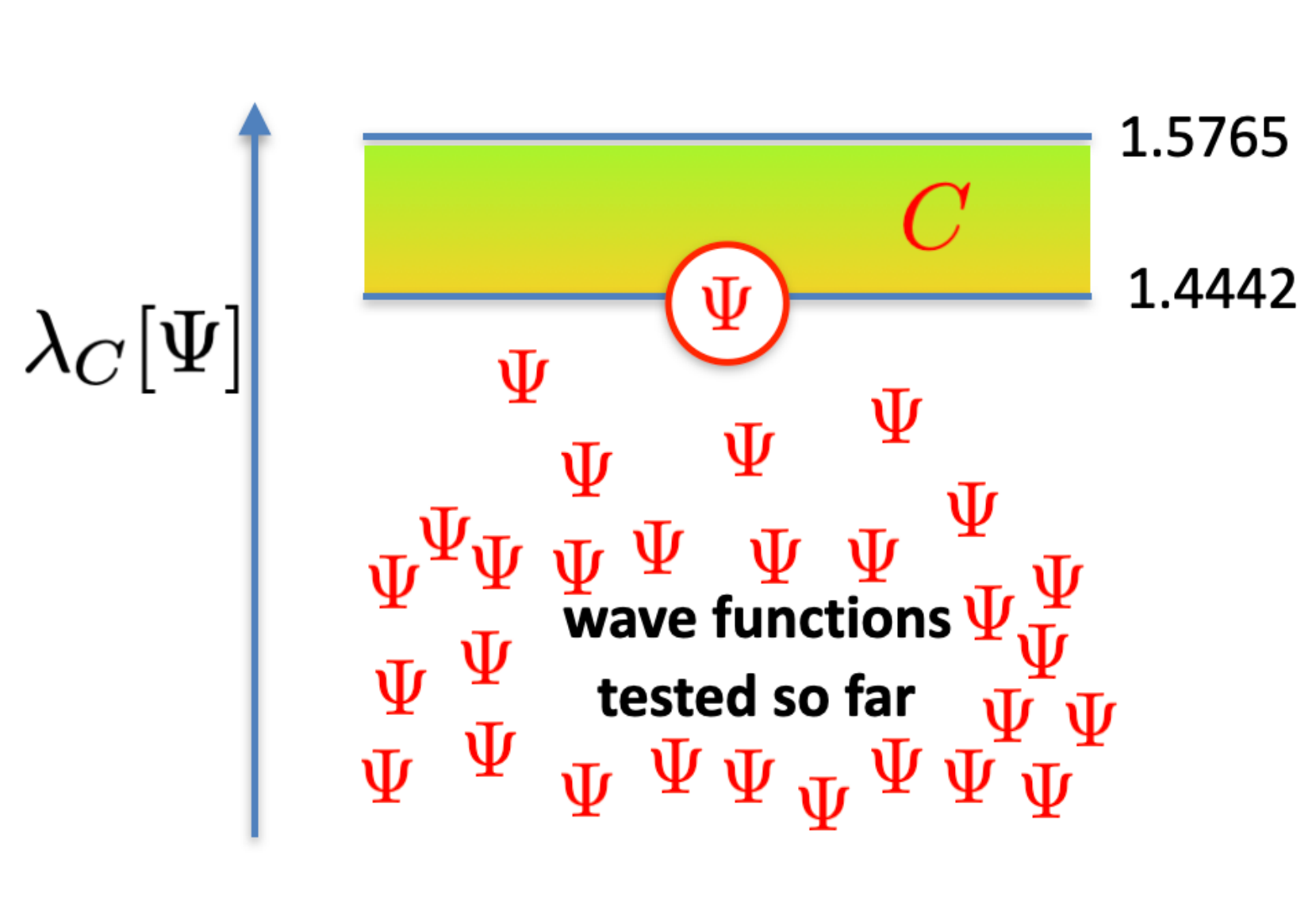}
\caption{Schematic illustration of the search for a lower bound for the optimal constant $C$ in the Lieb-Oxford inequality: each time we find (see the encircled $\Psi$ in the figure) a wavefunction (or just a probability density $|\Psi|^2$) yielding a value for $\lambda_C[\Psi]$ of Eq.~\eqref{LObound_Modify} higher than any value previously observed, we improve the lower bound for $C$. Currently, the highest value 1.4442 ever observed for $\lambda_C[\Psi]$ is provided by the bcc Wigner crystal energy \cite{LewLieSei-PRB-19,CotPet-arxiv-17}, but it is unknown whether this is the actual optimal value. The upper bounds on $C$, instead, are more difficult to derive, as they need a rigorous proof \cite{Lie-PLA-79,LieOxf-IJQC-81,ChaHan-PRA-99,LewLieSei-arxiv-22}. }
\label{fig:schematic}\end{figure}

We focus here on lower and upper bounds for the unknown optimal constants $C$ of Eq.~\eqref{CsupPsi} and $C(N)$ of Eq.~\eqref{eq:CNdef}, which can be obtained, respectively, by:
\begin{enumerate}
\item[$\bullet$] {\em Challenging} the LO bound of Eq.~\eqref{LObound_Original}, which means:\\
Finding a $\Psi$ with a particularly high value $\lambda_C[\Psi]$ in Eq.~\eqref{LObound_Modify}, yielding a {\em lower} bound $C\ge\lambda_C[\Psi]$.
\item[$\bullet$] {\em Tightening} the LO bound of Eq.~\eqref{LObound_Original}, which means:\\
Applying general theory to find an {\em upper} bound $\overline{C}$ for $C$, yielding $C\le\overline{C}$.
\end{enumerate}
This is illustrated schematically in Fig.~\ref{fig:schematic}, which shows the tightest upper and lower bounds currently known for $C$. The current best lower bound  corresponds to the $\Psi$ (or, better, the probability density $|\Psi|^2$), describing the bcc Wigner crystal \cite{LewLieSei-PRB-19,CotPet-arxiv-17}, which has been suggested several times \cite{Per-INC-91,RasPitCapPro-PRL-09} as the actual optimal $C$. However, this hypothesis still remains without proof.
The increasingly tighter lower and upper bounds for $C$ and $C(N)$ from the literature are also collected in table~\ref{tab:optC}.
\begin{table}[h]
\begin{tabular}{|l|l|l|}\hline
bound for $C$ and $C(N)$ & reference & year \\\hline\hline
$C\le 8.52$  & \cite{Lie-PLA-79}  & 1979 \\\hline
$C\le 1.68$  & \cite{LieOxf-IJQC-81} & 1981 \\\hline
$C\le 1.6359$  & \cite{ChaHan-PRA-99} & 1999 \\\hline
$C\le 1.5765$  & \cite{LewLieSei-arxiv-22} & 2022 \\\hline
$C\ge C(2)\ge 1.234\;=\;\Lambda_C[\rho_{81}]$ &  \cite{LieOxf-IJQC-81} & 1981 \\\hline
$C\ge C(2)\ge 1.256\;=\;\Lambda_C[\rho_{16A}]$ & \cite{SeiVucGor-MP-16}  & 2016 \\\hline 
$C\ge C(60)\ge 1.4119\;=\;\Lambda_C[\rho_{16B}]$ & \cite{SeiVucGor-MP-16}  & 2016 \\\hline 
$C\ge 1.4442$ & \cite{LewLieSei-PRB-19,CotPet-arxiv-17}  & 2019 \\\hline 
\end{tabular}
\caption{Upper and lower bounds for the optimal constants $C$  and $C(N)$ of Eq.~\eqref{CsupN},  where $\Lambda_C[\rho]$ is the density functional of Eq.~\eqref{Csuppho}.}\label{tab:optC}
\end{table}

The bound of Eq.~\eqref{LObound_Original} is not convex in $\rho$ (it is not even positive) \cite{Lie-IJQC-83}.
Alternatively, in Eq.~(5.8) of L83 Lieb defines the functional
\bmath
\tilde{I}(\rho)&=&\inf_{\Psi\mapsto\rho}\,\langle\Psi|\hat{V}_{\rm ee}|\Psi\rangle\nonumber\\
&\equiv&V_{\rm ee}^{\rm SCE}[\rho].
\label{DefVeeSCE}\emath
In the second line, we have introduced the designation $V_{\rm ee}^{\rm SCE}[\rho]$, Eq.~(21) of SVG \cite{SeiVucGor-MP-16}. There is no need to distinguish between pure states and ensembles in the definition of the functional \cite{Lew-CRM-18}. Furthermore, the minimization can be done over probability densities $P$ \cite{Lie-PLA-79,LieOxf-IJQC-81,CotFriKlu-CPAM-13},
\begin{equation}
V_{\rm ee}^{\rm SCE}[\rho] = \min_{P \mapsto \rho}\int \mathrm{d} \rv_1 \cdots\int \mathrm{d} \rv_N\,
P\big(\rv_1,\,\rv_2,\,...,\,\rv_N\big)\sum_{i > j} \frac{1}{|\rv_i - \rv_j|}.
\end{equation}
The SCE functional will be considered in detail in the following section \ref{secSCE}.
Instead of Eq.~\eqref{LObound_Original}, we now obtain a convex bound,
\bmath
\langle\Psi|\hat{V}_{\rm ee}|\Psi\rangle\;\ge\;\tilde{I}(\rho)\;\equiv\;V_{\rm ee}^{\rm SCE}[\rho],
\label{LObound_Alternative}\emath
see Eq.~(5.8) of L83. If the RHS $\tilde{I}(\rho)\equiv V_{\rm ee}^{\rm SCE}[\rho]$ had a simple explicit dependence on the density $\rho$, the bound given by Eq.~\eqref{LObound_Alternative} would be preferable to the one of Eq.~\eqref{LObound_Original}. In particular, Eqs.~\eqref{CsupPsi} and \eqref{eq:CNdef} can now be written \cite{SeiVucGor-MP-16} as a supremum with respect to densities $\rho$,
\bmath
C\;=\;\sup_{\rho}\,\Lambda_C[\rho],\qquad\qquad
C(N)\;=\;\sup_{\rho\,\mapsto N}\,\Lambda_C[\rho], \label{CsupN}
\emath
where the density functional $\Lambda_C[\rho]$ is defined as
\begin{equation}
    \Lambda_C[\rho]\;\equiv\;\frac{U[\rho]-V^{\rm SCE}_{\rm ee}[\rho]}{\int d^3r\,\rho(\rv)^{4/3}}. \label{Csuppho}
\end{equation}

\section{Strictly correlated electrons (SCE)}\label{secSCE}
We now turn in detail to the functional $V_{\rm ee}^{\rm SCE}[\rho]$, denoted as $\tilde{I}(\rho)$ in L83 and defined in Eq.~\eqref{DefVeeSCE} above. 

In Ref.~\cite{Lie-IJQC-83}, Elliott Lieb wrote:
{\em ``Any reader who is devoted to abstract density functional theory [...]
should try to guess a plausible form for $\tilde{I}(\rho)$. (Proving it is another matter.) It will quickly be seen that
$\tilde{I}(\rho)$ must be extremely complicated, and to say that it is ``nonlocal'' is an understatement.''}

By now, we know that the functional $\tilde{I}(\rho)$ corresponds to a multimarginal optimal transport (OT) problem \cite{ButDepGor-PRA-12,CotFriKlu-CPAM-13,FriGerGor-arxiv-22} and provides the semiclassical limit ($\hbar\to 0$) of the Levy-Lieb functional \cite{CotFriKlu-ARMA-18,Lew-CRM-18}.
But even before these rigorous works, the functional $\tilde{I}(\rho)$ was constructed explicitly from the physical idea of ``strictly-correlated electrons'' for some special cases: $N=2$ electrons in a spherically symmetric density \cite{Sei-PRA-99}, a general number of electrons $N$ in one dimensional systems \cite{Sei-PRA-99}, and, later, a general $N$-electron density with spherical symmetry \cite{SeiGorSav-PRA-07}. The first two constructions have been later proven to be exact in Refs.~\cite{ButDepGor-PRA-12} and \cite{ColDepDim-CJM-15}, respectively, while for the construction for the general spherical $N$-electron case counterexamples have been found \cite{ColStr-M3AS-15,SeiDiMGerNenGieGor-arxiv-17}. Nonetheless, the spherically-symmetric construction of Ref.~\cite{SeiGorSav-PRA-07} still provides a valid probability density corresponding to a given $\rho$, and thus a valid $\Lambda_C[\rho]$ that can be used to get improved lower bounds for the optimal constant \cite{RasSeiGor-PRB-11,SeiVucGor-MP-16}. 

In the following, we provide simple explicit constructions for $V_{\rm ee}^{\rm SCE}[\rho]$ for some of these special cases, and we then show that Lieb and Oxford had actually found in LO81, for $N=2$, a spherical density for which they could also construct what we call today the SCE (or Monge) solution.

\subsection{$N=2$ electrons with spherical symmetry (3D)}
We here consider spherically symmetric densities $\rho(\rv)=\rho(r)$ for $N=2$ electrons,
\bmath
\int_0^{\infty}dr\,(4\pi r^2)\,\rho(r)\;=\;N\qquad\qquad(N=2).
\label{densitySph2el}\emath
We shall first give the recipe for computing the value of $V_{\rm ee}^{\rm SCE}[\rho]\equiv\tilde{I}(\rho)$ for any given $\rho$, and then explain the physics behind it. The mathematical proof that $V_{\rm ee}^{\rm SCE}[\rho]$ for these densities yields the true infimum in Eq.~\eqref{DefVeeSCE} is given in Ref.~\cite{ButDepGor-PRA-12}.

{\bf Recipe:} (Although we consider the case $N=2$ here, we keep the symbol ``$N$'' in our equations to make the origin of the numerical prefactors transparent.) For a given density $\rho$, we introduce the cumulative particle number function,
\bmath
N_{\rm e}(r)\;=\;\int_0^r\di s\,(4\pi s^2)\,\rho(s).
\label{defNe}\emath
As $r\ge0$ grows, $N_{\rm e}(r)$ increases monotonically from $N_{\rm e}(0)=0$ to $N_{\rm e}(r)\to N$ for $r\to\infty$.
When $\rho$ has compact support, $\rho(r)=0$ for $r>R$, we have $N_{\rm e}(R)=N$. Generally, including the cases with $R=\infty$, we use for the inverse function the notation
\bmath
N^{-1}_{\rm e}:\;[0,N]\to[0,R],\qquad\nu\mapsto r=N^{-1}_{\rm e}(\nu).
\emath
Then, fixed by the density $\rho(r)$, the SCE ``co-motion function'' $f(r)$ (aka, the Monge map in optimal transport theory) is defined as
\bmath
f(r)\;=\;N_{\rm e}^{-1}\big(2\,-\,N_{\rm e}(r)\big),
\label{fNe}\emath
and the value of the density functional $V_{\rm ee}^{\rm SCE}[\rho]\equiv\tilde{I}(\rho)$ is given by
\bmath
V_{\rm ee}^{\rm SCE}[\rho]\;=\;\int_0^\infty dr\,(4\pi r^2)\,\frac{\frac1N\,\rho(r)}{r+f(r)}\qquad\qquad(N=2).
\label{VeeSCEspher2}\emath
This is indeed a highly nonlocal functional of the density $\rho$~!


{\bf Physics:} In an SCE state with a density according to Eq.~\eqref{densitySph2el}, minimum repulsion energy in 
Eq.~\eqref{DefVeeSCE} is achieved as follows: The two electrons (upon simultaneous measurement of their positions)
are always found at positions $\rv_1$ and $\rv_2$ which are related by
\bmath
\rv_2\;=\;\fv(\rv_1),\qquad\fv(\rv)\;=\;-f(r)\,\frac{\rv}{r}.
\emath
In words: The vectors $\rv_1$ and $\rv_2$ always point in opposite directions from the origin, and the distance $r_2=|\rv_2|$ is fixed by $r_1=|\rv_1|$ through the co-motion function, $r_2=f(r_1)$.
Formally, using a Dirac $\delta$ distribution, the square modulus of such an SCE state can be written as
\bmath
|\Psi(\rv_1,\rv_2)|^2\;=\;\frac{\rho(\rv_1)}{N}\,\delta\big(\rv_2-\fv(\rv_1)\big).
\emath
Eq.~\eqref{fNe} arises, since the probability for finding one electron inside a sphere with radius $r$ must be the same as for finding the second one outside a sphere with radius $f(r)$,
\bmath
N_{\rm e}(r)\;=\;2\,-\,N_{\rm e}\big(f(r)\big).
\emath
Then Eq.~\eqref{VeeSCEspher2} arises, since the distance $d(r)=r+f(r)$ between the two electrons only depends on the single variable $r$ (representing the distance $r=|\rv|$  of one of them from the center). 
Eq.~\eqref{fNe} implies that $f(f(r))\equiv r$, meaning that $f$ must be its own inverse,
\bmath
f^{-1}(r)\;=\;f(r).
\emath
This result is necessary for identical particles, in which case the above condition $r_2=f(r_1)$ must be equivalent to $r_1=f(r_2)$.

\begin{figure}[htb]
\includegraphics[width=80mm]{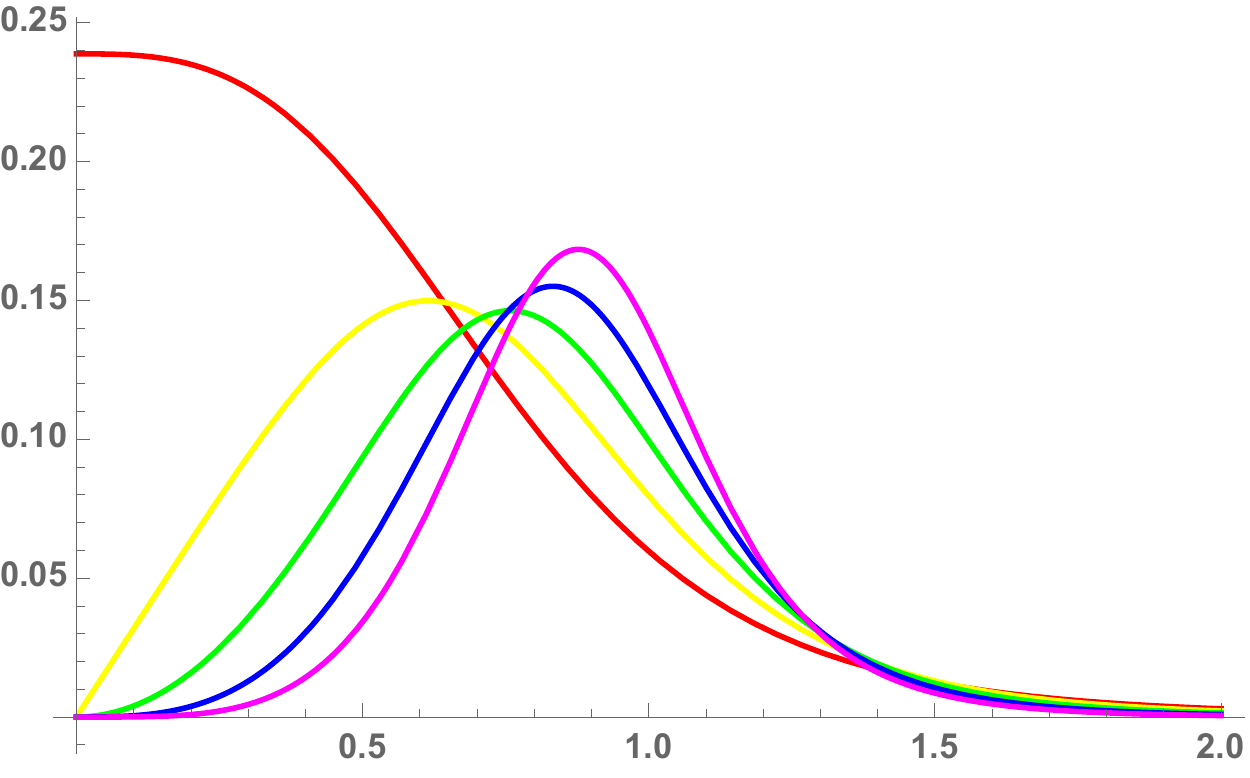}
\centering
\put(-290, 120){\Large$\frac{a^3}N\,\rho_n(r)$} \put(10, 5){\Large$r/a$} 
\caption{Densities $\rho_n(r)$ of Eq.~\eqref{rhoExpl1} for $n=3$ (red), 4 (yellow), ..., up to $n=7$ (violet).}
\label{figLO21_1}\end{figure}

{\bf Example 1:} For illustration, we choose the cumulative number functions
\bmath
N_{\rm e}(r)\;=\;\frac{Nr^n}{a^n+r^n}\qquad\qquad\qquad(a,n>0,\quad N\in\N)
\label{explNe1a}\emath
(later we shall set $N=2$). These functions can be inverted explicitly,
\bmath
N^{-1}_{\rm e}(\nu)\;=\;a\left(\frac{\nu}{N-\nu}\right)^{1/n}\qquad\qquad(a,n>0,\quad N\in\N).
\emath
In terms of the derivatives $N'_{\rm e}(r)=\frac{d}{dr}\,N_{\rm e}(r)$, the corresponding densities are (see Fig.~\ref{figLO21_1})
\bmath
\rho_n(r)\;\equiv\;\frac{N'_{\rm e}(r)}{4\pi r^2}\;=\;\frac{N}{4\pi a^3}\,\frac{nx^{n-3}}{(1+x^n)^2}\qquad\left(x=\frac{r}a\right).
\label{rhoExpl1}\emath
%
Setting $N=2$, we obtain a unique co-motion function $f(r)$, valid for all $n$,
\bmath
f(r)\;\equiv\;N_{\rm e}^{-1}\big(N\,-\,N_{\rm e}(r)\big)\;=\;a\left(\frac{N-N_{\rm e}(r)}{N_{\rm e}(r)}\right)^{1/n}\;=\;\frac{a^2}r\qquad\qquad(N=2).
\emath
Consequently, Eq.~\eqref{VeeSCEspher2} now yields
\bmath
V_{\rm ee}^{\rm SCE}[\rho_n]
&\equiv& \frac1N\int_0^\infty dr\,\frac{N'_{\rm e}(r)}{r+f(r)}\qquad\qquad(N=2)\label{VeeSCE_Ne}\\
&=&      na^n\int_0^\infty\frac{dr\,r^n}{(a^2+r^2)(a^n+r^n)^2}\;=\;\frac1a\,V_n,
\emath
with the dimensionless integrals
\bmath
V_n\;=\;n\int_0^\infty\frac{dx\,x^n}{(1+x^2)(1+x^n)^2}.
\emath
For spherical densities $\;\rho(r)=\frac{N'_{\rm e}(r)}{4\pi r^2}$, \;we generally have
\bmath
U[\rho]&\equiv&(4\pi)^2\int_0^R\di r\,r\,\rho(r)\int_0^r\di s\,s^2\,\rho(s)\nonumber\\
&=&\int_0^R\frac{\di r}r\,N'_{\rm e}(r)\,N_{\rm e}(r),\label{UNe}\\
\int d^3r\,\rho(\rv)^{4/3}&=&\frac{1}{(4\pi)^{1/3}}\int_0^R\frac{\di r}{r^{2/3}}\,N'_{\rm e}(r)^{4/3}.
\label{ExNe}\emath
Using here the functions $N_{\rm e}(r)$ from Eq.~\eqref{explNe1a}, we obtain (for arbitrary $N\in\N$)
\bmath
U[\rho_n]&=&\;N^2na^n\int_0^{\infty}\frac{dr\,r^{2(n-1)}}{(a^n+r^n)^3}\qquad\;=\quad\frac{N^2}{a}\,U_n,\\
\int d^3r\,\rho_n(\rv)^{4/3}&=&\frac{(Nna^n)^{4/3}}{(4\pi)^{1/3}}\int_0^{\infty}\frac{dr\,r^{\frac{4n}3-2}}{(a^n+r^n)^{8/3}}
\;=\;\frac{N^{4/3}}{a}\,K_n,
\emath
with another set of dimensionless integrals,
\bmath
U_n\;=\;n\int_0^{\infty}\frac{dx\,x^{2(n-1)}}{(1+x^n)^3},\qquad\qquad
K_n\;=\;\frac{n^{4/3}}{(4\pi)^{1/3}}\int_0^{\infty}\frac{dx\,x^{\frac{4n}3-2}}{(1+x^n)^{8/3}}.
\emath
For $N=2$, we now can evaluate the density functional of Eq.~\eqref{Csuppho},
\bmath
\Lambda_C[\rho_n]\;=\;\frac{2^2U_n-V_n}{2^{4/3}K_n}.
\emath
The resulting values for the densities shown in Fig.~\ref{figLO21_1} are collected in the Table below.

\begin{tabular}{|c|c|}\hline
$n$ & $\Lambda_C[\rho_n]$ \\\hline\hline
3 & 1.252~22 \\\hline 
4 & 1.234~16 \\\hline 
5 & 1.205~83 \\\hline 
6 & 1.175~75 \\\hline 
7 & 1.146~48 \\\hline 
\end{tabular}

For many integer values of $n$, the integrals $V_n$, $U_n$, and $K_n$ can be obtained analytically. 
Numerically, the maximum seems to be reached for $n\approx2.86$, and it is equal to $\Lambda_C[\rho_n]\approx1.252~62$. This value is only slightly lower than the currently best known lower bound on the Lieb-Oxford constant $C(2)$, namely 1.256 \cite{SeiVucGor-MP-16}, and it is significantly better than the original LO81 lower bound, equal to 1.234, discussed in the next section.

\subsection{An early example for an SCE state in LO81} \label{sec:LOSCE}
In section 3 of LO81, Lieb and Oxford introduce for $N=2$ electrons in 3D the spherically symmetric probability density $|\psi|^2[(t,e),(s,h)]$, which we denote $|\Psi_{81}|^2$ here. The authors point out that {\em ``trivially, $I(\Psi_{81})\equiv\langle\Psi_{81}|\hat{V}_{\rm ee}|\Psi_{81}\rangle=1$ since the particles are always one unit apart"}, to find for the present Eq.~\eqref{LObound_Modify}
\bmath
C(2)\;\ge\;\lambda_C[\Psi_{81}]\;=\;1.234.
\emath
$|\Psi_{81}(\rv_1,\rv_2)|^2$ turns out to describe an SCE state for $N=2$ electrons in the spherical density (LO81 uses dimensionless lengths in units of our $R$ here)
\bmath
\rho_{81}(r)\;=\;\frac{15}{\pi R^5}\,(R-r)^2\itTh(R-r).
\label{rho81}\emath
In this case, Eq.~\eqref{fNe} implies (see derivation below) the simple SCE co-motion function,
\bmath
f(r)\;=\;R-r,
\label{co-motionF81}\emath
enforcing in fact a fixed distance $r+f(r)=R$ between the two electrons ({\em ``the particles are always one unit apart''}),
\bmath
V^{\rm SCE}_{\rm ee}[\rho_{81}]\;=\;\frac1R.
\label{VeeSCE81}\emath
To show that $f(r)=R-r$ is the true co-motion function for the density $\rho_{81}(r)$, we study the set of all possible cumulative number functions $\;N_{\rm e}:[0,R]\to[0,2],\;r\mapsto N_{\rm e}(r)$, \;that in Eq.~\eqref{fNe} yield $f(r)=R-r$,
\bmath
N_{\rm e}(R-r)\;=\;2\,-\,N_{\rm e}(r).
\label{co-motionFproof}\emath
\begin{enumerate}
\item[$\bullet$] Specifically for the density $\rho_{81}(r)$, with
\bmath
N_{\rm e}(r)\;\equiv\;\int_0^r\di s\,(4\pi s^2)\,\rho_{81}(s)\;=\;\frac2{R^5}\,\Big[6r^2\,-\,15R\,r\,+\,10R^2\Big]\,r^3,
\label{Ne81}\emath
we see that condition \eqref{co-motionFproof} is satisfied, by replacing $r$ with $R-r$ and expanding.
There is no need for explicitly constructing the inverse function $N_{\rm e}^{-1}(\nu)$ here.
\item[$\bullet$] For the general case, we substitute $r=\frac{R}2+u$ to rewrite condition \eqref{co-motionFproof} as
\bmath
\textstyle N_{\rm e}\big(\frac{R}2-u\big)\;=\;2\,-\,N_{\rm e}\big(\frac{R}2+u\big).
\label{invSymm}\emath
In words: The co-motion function $f(r)=R-r$ arises then and only then, when the graph $y=N_{\rm e}(x)$ in the $xy$-plane has {\bf inversion symmetry} about the point $(x,y)=(\frac{R}2,1)$.
\end{enumerate}
Using Eq.~\eqref{invSymm}, we can construct infinitely many densities $\rho(r)=\frac{N'_{\rm e}(r)}{4\pi r^2}$ (for $r\in[0,R]$)
that all share the simple co-motion function $f(r)=R-r$ of Eq.~\eqref{co-motionF81}, implying that
\bmath
V^{\rm SCE}_{\rm ee}[\rho]\;=\;\frac1R.
\label{VeeSCE81gen}\emath

{\bf Example 2:} For $-\frac32\le a\le\frac38$, all the functions
\bmath
y_a(s)\;=\;a\,s^5\,-\,\big(\textstyle{2a+\frac12}\big)\,s^3\,+\,\big(\textstyle{a+\frac32}\big)\,s\qquad\qquad
\big(\!-\frac32\le a\le\frac38\big),
\emath
having inversion symmetry about the origin $(s,y)=(0,0)$, are monotonic for $s\in[-1,1]$ with $y_a(\pm1)=\pm1$ [and $y'_a(\pm1)=0$].
Consequently, Eq.~\eqref{invSymm} is satisfied when we choose
\bmath
N_{\rm e}(r)&=&1\,+\,y_a\big({\textstyle\frac{2r-R}R}\big)\nonumber\\
&=&32\,a\Big(\frac{r}R\Big)^5-80\,a\Big(\frac{r}R\Big)^4+\,4\,\big(16a-1\big)\Big(\frac{r}R\Big)^3+\,2\,\big(3-8a\big)\Big(\frac{r}R\Big)^2.
\label{NeExample}\emath
The resulting densities $\rho_a(r)=\frac{N'_{\rm e}(r)}{4\pi r^2}$ all have the simple co-motion function $f(r)=R-r$.
With $a=\frac38$, the density $\rho_{81}(r)$ of LO81, Eq.~\eqref{rho81}, is included, see Eq.~\eqref{Ne81}. 

\begin{figure}[htb]
\includegraphics[width=80mm]{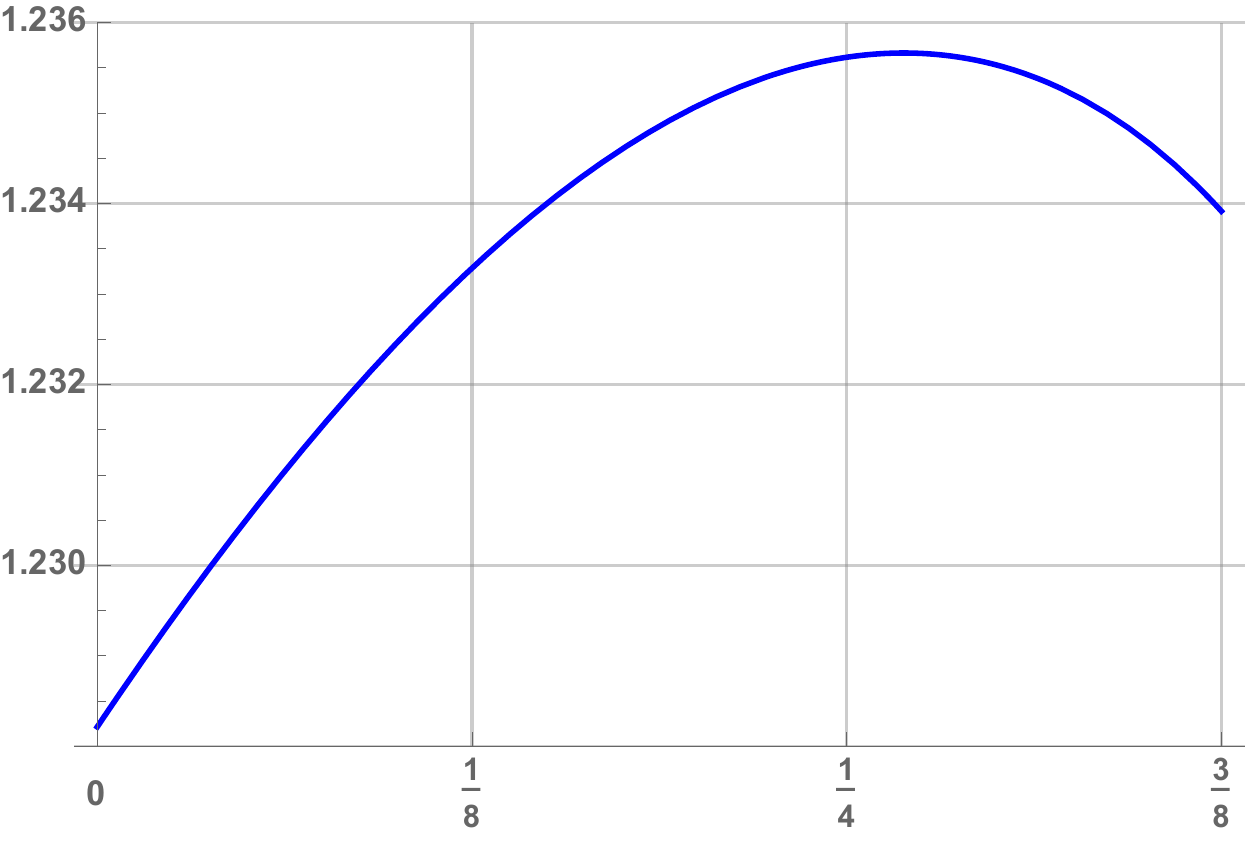}
\centering
\put(-280, 130){\Large$\Lambda_C[\rho_a]$} \put(5, 13){\Large$a$} 
\caption{The values $\Lambda_C[\rho_a]$ for the functions $N_{\rm e}(r)$ of Eq.~\eqref{NeExample}.
}
\label{figLO21_4}\end{figure}

Using Eqs.~\eqref{UNe}, \eqref{ExNe}, plus Eq.~\eqref{VeeSCE81gen}, we can evaluate the functional  $\Lambda_C[\rho_a]$ of Eq.~\eqref{Csuppho} for different values of $a\in[-\frac32,+\frac38]$, see Fig.~\ref{figLO21_4}.
We see that the maximum $\Lambda_C\approx1.2358$ is reached at $a\approx0.27<\frac38$, slightly higher than the value $\Lambda_C=1.234$ reported in LO81 for $a=\frac38$.

\section{SCE for a density with two ``bumps'' in 1D}
\label{secBumps}
Again in L83, Elliott Lieb elaborates on the extreme nonlocality of $\tilde{I}(\rho)$ and writes {\em``consider $N=2$ and $\rho$ consisting of two ``bumps'' $\rho_1$ and $\rho_2$, very far apart. As long as $\int\rho_1=\int\rho_2=1$, $\tilde{I}(\rho)\approx 0$, independently of $\rho_1$ and $\rho_2$. But when $\int\rho_1>1,\;\int\rho_2<1$, then $\tilde{I}(\rho)$ depends heavily on $\rho_1$ but not on $\rho_2$. The reason is that in the former case the two electrons can be far apart in the two bumps; in the latter case the two electrons must be partly close together in the first bump.''}\footnote{We have corrected here a small typo in this sentence}

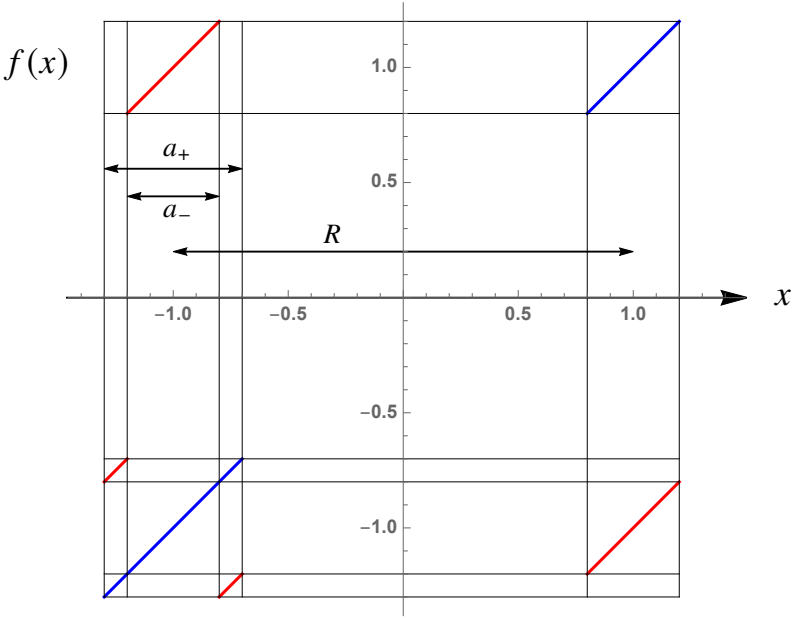
\begin{figure}[htb]
\includegraphics[width=90mm]{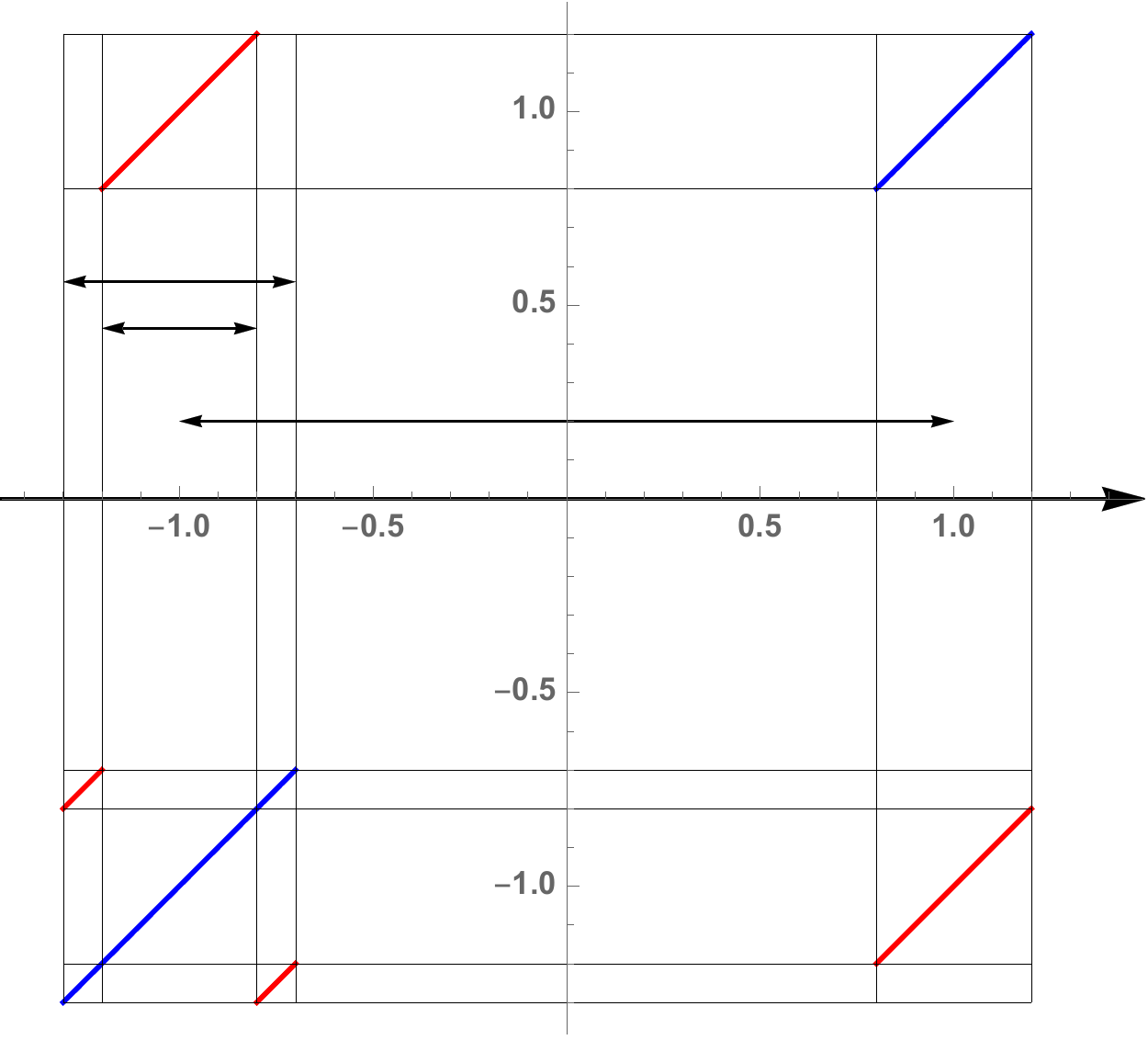}
\centering
\put(-280, 205){\Large$f(x)$} \put(10, 118){\Large$x$}
\put(-220, 174){$a_+$} \put(-220, 151){$a_-$} \put(-160, 141){$R$}
\caption{Co-motion functions $f(x)$ (red) and  $x$ (blue)
 in the case $R=2$, $a=0.5$, and $\de=0.2$ (when $a_+=0.6$ and $a_-=0.4$). Generally, $f$ is its own inverse, $f^{-1}(x)=f(x)$.}
\label{figLO21_2}\end{figure}

Since we know now how to exactly construct $\tilde{I}(\rho)$ for one-dimensional systems \cite{ColDepDim-CJM-15}, we follow here this suggestion, and consider a density $\rho_{\de}(x)$ for $N=2$ electrons in 1D, consisting of two rectangular ``bumps'' with different widths $a_{\pm}=(1\pm\de)a$ (but with equal local densities $\rho_0=\frac1a$), separated center-to-center by a large distance $R\gg a$,
\bmath
\rho_{\de}(x)\;=\;\left\{\begin{array}{cc}
\frac1a & \Big(|x+\frac{R}2|\le\frac{a_+}2\Big) \\ \frac1a & \Big(|x-\frac{R}2|\le\frac{a_-}2\Big) \\
     0      & \Big(\text{elsewhere}\Big)
\end{array}\right\}\qquad\Big(a_{\pm}=(1\pm\de)a,\quad 0\le\de\le1\Big).
\label{rho-bumps}\emath
In this 1D case without ``radial'' symmetry, $\rho(-x)\ne\rho(x)$, the analogue of Eq.~\eqref{fNe} reads
\bmath
f(x)\;=\;\left\{\begin{array}{cc}
N_{\rm e}^{-1}\big(N_{\rm e}(x)+1\big) & \Big(N_{\rm e}(x)<1\Big),\\
N_{\rm e}^{-1}\big(N_{\rm e}(x)-1\big) & \Big(N_{\rm e}(x)>1\Big),\end{array}\right.
\label{fNe1D}\emath
with the cumulative particle number function $N_{\rm e}(x)=\int_{-\infty}^x\di y\,\rho(y)$ and its inverse $N^{-1}_{\rm e}(\nu)$.
Eq.~\eqref{fNe1D} arises, as two strictly correlated electrons in 1D are always separated by a distance over which the density integrates exactly to one particle,
\bmath
\int_x^{f(x)}\di x\,\rho(x)\;=\;\left\{\begin{array}{cc} 1 & \Big(N_{\rm e}(x)<1\Big), \\
                                                        -1 & \Big(N_{\rm e}(x)>1\Big).\end{array}\right.
\emath
For the density $\rho_{\de}(x)$, the function $f(x)$ is plotted in Fig.~\ref{figLO21_2} (red linear segments).

Fig.~\ref{figLO21_2} shows that the distance between these two electrons is always either $a$ or $R$,
\bmath
|f(x)-x|\;=\;\left\{\begin{array}{cc}
a & (x_1<x<x_2) \\ R & (x_2<x<x_3 )\\ a & (x_3<x<x_4) \\ R & (x_5<x<x_6)
\end{array}\right\},
\emath
where $x_{1,4}=-\frac{R}2\mp\frac{a_+}2$, $x_{2,3}=-\frac{R}2\mp\frac{a_-}2$, $x_{5,6}=+\frac{R}2\mp\frac{a_-}2$.
Consequently (with $N=2$),
\bmath
V^{\rm SCE}_{\rm ee}[\rho_{\de}]&\equiv&\int_{-\infty}^{\infty}\frac{\frac1N\,\rho_{\de}(x)\,\di x}{|f(x)-x|}\nonumber\\
&=&\frac{\rho_0}{N}\left[\frac{x_2-x_1}a+\frac{x_3-x_2}R+\frac{x_4-x_3}a+\frac{x_6-x_5}R\right] \nonumber \\
&=& \frac{\de}a\,+\,\frac{1-\de}R.
\emath
As was to be expected, we find $V^{\rm SCE}_{\rm ee}[\rho_{0}]=\frac1R$ for $\de=0$ (with symmetric bumps), as this is the only case when each bump always accommodates exactly one electron, implying that the distance between these strictly correlated electrons is always exacly $R$. In the opposite case $\de=1$,
when $a_+=2a$ and $a_-=0$, there is only one single bump with width $2a$, and the two electrons always have the fixed distance $a$. Then, as predicted by Lieb, if we take the limit $R\to \infty$ we see that for $\delta=0$ the functional is zero, but when $\delta > 0$ we have a dependence on $\rho_1$ via the parameter $a$.

\section{Which densities are the most challenging for the bound?}
We conclude this chapter with an intriguing question: since we can now turn the search for improved lower bounds for $C(N)$ from wavefunctions to densities via the functional $\Lambda_C[\rho]$ of Eq.~\eqref{Csuppho}, which kind of densities provide higher values of $\Lambda_C[\rho]$ for a given number of electrons $N$?

In Ref.~\cite{SeiVucGor-MP-16} several different density profiles for $N=2$ electrons in a spherically symmetric density were used to compute $\Lambda_C[\rho]$. We report some of them, together with a few new ones, in table~\ref{tab:valuesdens}. We observe that an exponential density gives a rather high value of $\Lambda_C[\rho]$ and the profile $\sqrt{r}\, e^{-r}$ even more so. Here we have also extended this investigation to $N>2$, by using the variational solution of Ref.~\cite{SeiGorSav-PRA-07} for spherically symmetric densities. For example,
in table~\ref{tab:valuesdensN3} we have played with spherical density profiles of various atoms and rescaled them such that they integrate to 3. We were surprised to see that, again, the exponential profile (H atom) and the profile $\sqrt{r}\, e^{-r}$ give particularly high values. The same happens for $N=4$ and $N=10$, as shown in table~\ref{tab:valuesdensN10}. 

\begin{table}[h]
\begin{tabular}{|l|l|}\hline
profile & $\Lambda_{C}[\rho]\; (N=2)$ \\ \hline\hline
$e^{-50\,(r-1)^2}$  & $0.932$  \\ \hline
$\theta(1-r)$  & $1.106$  \\ \hline
$(1+r)^{-4}$  & $1.154$  \\ \hline
$e^{-\sqrt{r}}$ & $1.224$\\ \hline
$e^{-r^{3/2}}$ & $1.254$\\ \hline
$e^{-r}$ & $1.255$ \\ \hline
$\sqrt{r}\,e^{-r}$ & $1.256$ \\ \hline
\end{tabular}
\caption{Values of the functional $\Lambda_{C}[\rho]$ for different spherically-symmetric density profiles for the case $N=2$. The function $\theta(x)$ is the Heaviside step function.}\label{tab:valuesdens}
\end{table}

\begin{table}[h]
\begin{tabular}{|l|l|}\hline
profile & $\Lambda_{C}[\rho]\; (N=3)$ \\ \hline\hline
$\theta(1-r)$  & $1.145$  \\ \hline
$\rho_{\rm B}(r)$  & $1.211$  \\ \hline
$\rho_{\rm Be}(r)$  & $1.235$  \\ \hline
$\rho_{\rm Li}(r)$  & $1.265$  \\ \hline
$\rho_{\rm H}(r)$ & $1.279$\\ \hline
$\sqrt{r}\,e^{-r}$ & $1.282$ \\ \hline
\end{tabular}
\caption{Values of the functional $\Lambda_{C}[\rho]$ for different spherically-symmetric density profiles for the case $N=3$. As an experiment, we have taken radial densities of various atoms (B, Be, H) and rescaled them such that they integrate to $N=3$. The radial densities for B, Be and H were computed using full CI in an aug-cc-pVDZ basis using pyscf \cite{Sun-WIRCMS-18}. The function $\theta(x)$ is the Heaviside step function. }\label{tab:valuesdensN3}
\end{table}

\begin{table}[h]
\begin{tabular}{|l|l|l|}\hline
profile & $\Lambda_{C}[\rho]\; (N=4)$ & $\Lambda_{C}[\rho]\; (N=10)$ \\ \hline\hline
$\theta(1-r)$  & $1.184$ &  $1.261$ \\ \hline
$\rho_{Z=N}(r)$  & $1.278$ &  $1.340$ \\ \hline
$e^{-r}$ & $1.307$ & $1.364$ \\ \hline
$\sqrt{r}\,e^{-r}$ & $1.310$ & $1.368$ \\ \hline
\end{tabular}
\caption{Values of the functional $\Lambda_{C}[\rho]$ for different spherically-symmetric density profiles for the cases $N=4$ and $N=10$, including the corresponding neutral atom densities computed using full CI in an aug-cc-pVDZ basis using pyscf \cite{Sun-WIRCMS-18}. The function $\theta(x)$ is the Heaviside step function.}\label{tab:valuesdensN10}
\end{table}



\begin{funding}
This work was supported by the Netherlands Organisation for Scientific Research (NWO) under Vici grant 724.017.001. T.B. is grateful to the Vrije Universiteit for the opportunity to contribute to this paper using the University Research Fellowship.
\end{funding}


\bibliographystyle{unsrtnat}
\bibliography{biblio_all.bib}









\end{document}